\documentclass[traditabstract,a4,letter]{aa}  

\usepackage{epsfig}
\usepackage{graphicx}
\usepackage{epstopdf}
\usepackage{color}  
\usepackage{array}   
\usepackage{units}  
\usepackage[breaklinks,colorlinks, citecolor=blue]{hyperref}
\usepackage{natbib}  
\usepackage{multirow}   
\usepackage{amsmath,amssymb}  
\usepackage[english]{babel}
\usepackage[latin1]{inputenc}
\usepackage[T1]{fontenc}
\usepackage{longtable,lscape}
\usepackage{footnote}

\usepackage{txfonts}
\usepackage[toc,page]{appendix} 
\begin{document}

%%%%%%%%%%%%%

%%%%%%%%%%%%%%%
\title{First measurement of the cross-correlation between\\ CMB weak lensing and X-ray emission}
\author{G.Hurier\inst{1}, P. Singh\inst{2}, C. Hern\'andez-Monteagudo\inst{1}}

\institute{
1 Centro de Estudios de F\'isica del Cosmos de Arag\'on (CEFCA),Plaza de San Juan, 1, planta 2, E-44001, Teruel, Spain\\
2 Inter-University Centre for Astronomy and Astrophysics, Ganeshkhind, Post Bag 4, Pune 411007, India\\
\\
\email{ghurier@ias.u-psud.fr} 
}

\abstract{Since the publication of the results of the {\it Planck} satellite mission in 2013, the local and early universes have been considered to be in tension in respect of the determination of amplitude of the matter density spatial fluctuations ($\sigma_8$) and the amount of matter present in the universe ($\Omega_m$). This tension can be seen as a lack of massive galaxy clusters in the local universe compared to the prediction inferred from {\it Planck} cosmic microwave background (CMB) best-fitting cosmology. In the present analysis, we perform the first detection of the cross-correlation between X-rays and CMB weak-lensing at 9.1\,$\sigma$. We next combine thermal Sunyaev-Zel'dovich, X-rays, and weak-lensing angular auto and cross power spectra to determine the galaxy cluster hydrostatic mass bias. We derive $(1-b_H) = 0.70 \pm 0.05$. Considering these constraints, we observe that estimations of $\sigma_8$ in the local Universe are consistent with {\it Planck} CMB best-fitting cosmology. However, these results are in clear tension with the output of hydrodynamical simulations that favor $(1-b_H) > 0.8$.}


\authorrunning{G.Hurier et al.}
\titlerunning{First measurement of the cross-correlation between CMB weak lensing and X-ray emission }

\maketitle
  
\section{Introduction}

Modern cosmology relies heavily in the observations and analysis of the cosmic microwave background (CMB) data. If the standard six parameter $\Lambda$CDM model provides a satisfying description of the outcome of the main cosmological probes \citep{planckcmb}, the determination of its parameters appears to be tension when separately considering the CMB angular power spectrum on the one hand, and the abundance of galaxy cluster in the local universe ($z < 1$) on the other \citep[see e.g.,][for a recent CMB-galaxy cluster joint analysis]{sal17}.

One current challenge of modern cosmology is thus related to the understanding of the origin of this apparent tension. This could be produced by our lack of knowledge regarding the galaxy cluster mass-observable relations, or by new physics beyond the standard $\Lambda$CDM model affecting the structure growth between the CMB last scattering surface and the local universe \citep{sal17,planckSZC}.
In this context, several probes can be used to trace the large scale distribution of matter in the universe. During their propagation along the line of sight, the CMB photons are affected by several physical processes such as the inverse Compton scattering expressed by the thermal Sunyaev-Zel'dovich effect \citep[hereafter tSZ,][]{sun69,sun72}, and the CMB gravitational lensing \citep{bla87} accounting for the deflections induced by the gravitational potential integrated along the line of sight, $\phi$.
The hot electrons causing the tSZ effect are also emitting in the X-ray domain through Bremsstrahlung radiation, in such a way that 
%At microwave frequencies, additionnal foreground emissions contribute to the total signal of the sky, for example the cosmic infra-red background \citep[CIB,][]{pug96,fix98}.\\
the tSZ effect, X-ray emission, and gravitational lensing have been powerful sources of cosmological and astrophysical constraints \citep[see, e.g.,][]{planckSZC,planckphi}.

However, the use of galaxy clusters as cosmological probes often requires the determination of their total mass, a complicated step usually relying on the hydrostatic equilibrium hypothesis. While such mass estimates are known to be biased, $M_{\rm hydro}/M_{\rm true} = 1-b_H$, hydrodynamical simulations favor low values for this bias, $b_H < 0.2$ \citep{lau13,bif16}. At the same time, on the observational side a significant amount of analyses combining X-ray, tSZ and weak lensing observations have been conducted in the last five years, either based on an object by object approach, or resorting to stacking algorithms \citep[see, e.g.,][]{hur17c,med17,ser17,jim17,par17,oka16,bat16,app16,smi16,hoe15,sim15,isr15,van14,don14,gru14,mah13}.
These analyses obtain an hydrostatic mass bias in the range $b_H = 0.20 \pm 0.08$, which seem to be compatible with the measurement obtained after combining CMB weak lensing and tSZ measurements towards SDSS DR8 redMaPPer \citep{ryk14} galaxy clusters, $b_H = 0.26 \pm 0.07$ \citep{hur17c} .

The statistical correlations of these tracers on the sky have also driven a lot of attention. For instance, the correlation between tSZ-X and tSZ-$\phi$ cross-analyses have been used to set cosmological constraints \citep[see,  e.g.,][]{hil14,hur15a,hur15b}. These measurements present different sensitivities to cosmological parameters and mass-observable relations. Consequently, a coherent statistical analysis of $\phi$, tSZ, and X-ray emission is a powerful tool to identify the origin of the tension between the CMB and the low redshift universe.

In the present paper, we present the first measurement of the CMB weak lensing and X-ray emission cross-correlation power spectrum. We model this cross-correlation using a halo-model formalism to derive cosmological constraints. Finally, we combine this result with previous studies to derive constraints on the hydrostatic mass bias.

%%%%%

\section{Modelling the tSZ, X-ray, and $\phi$ cross-correlations}
\label{crossth}

\begin{figure*}[!h]
\begin{center}
\includegraphics[width=0.9\linewidth]{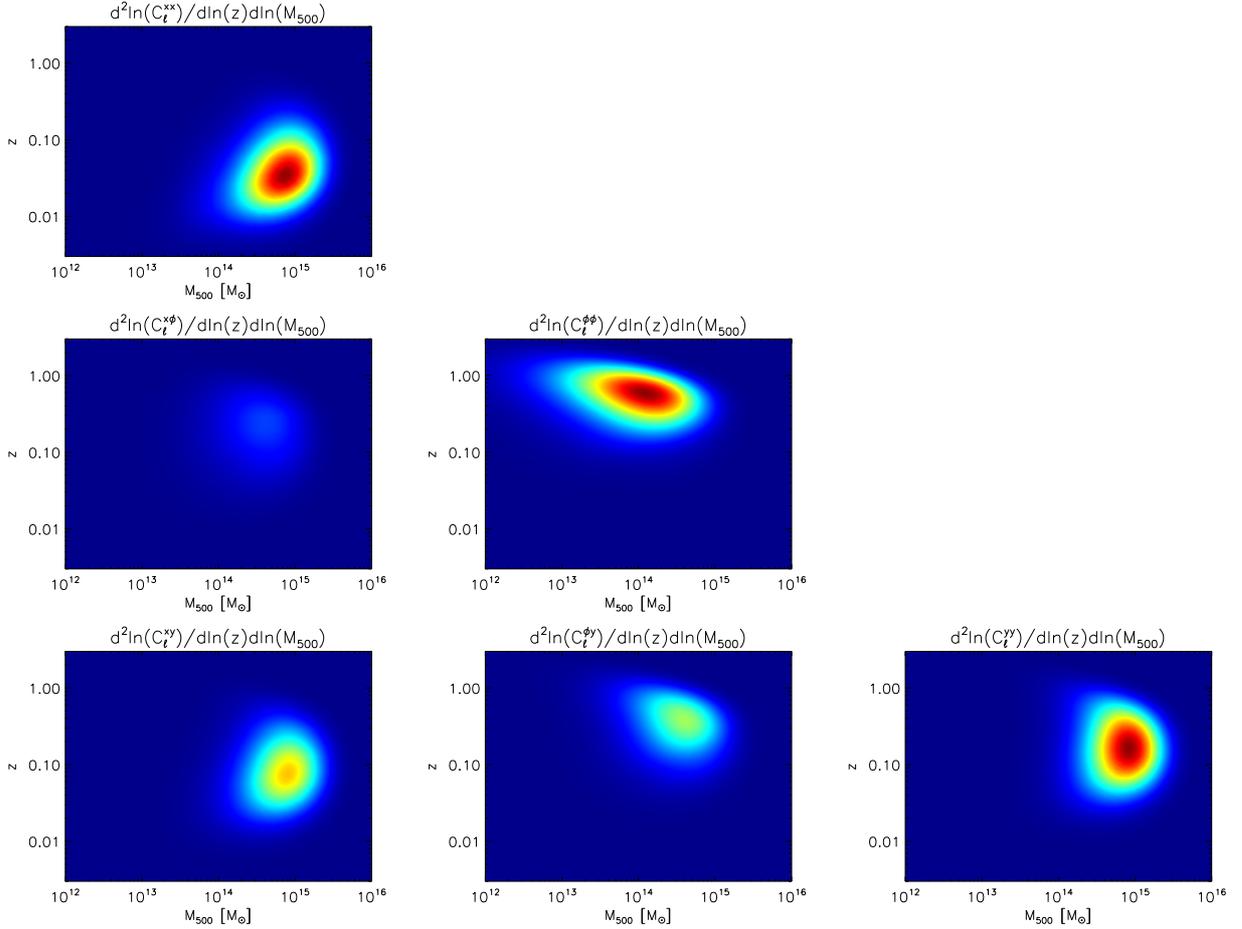}
\caption{Power density for the one-halo term, $\frac{{\rm d}^2\ln{\rm (C_\ell)}}{{\rm d}{\rm \ln}(M_{500}){\rm d}\ln(z)}$, as a function of halo mass, $M_{500}$, and redshift, $z$ at $\ell = 200$. On the diagonal we display power density for the auto-correlation power spectra. From left to right: contribution to X-ray, CMB weak lensing, and tSZ angular power spectrum at $\ell=200$. Off-diagonal panels represent power density of cross-power spectra, from left to right and top to bottom referring to: X-ray--lensing, X-ray--tSZ, and lensing--tSZ cross power spectra. The color scale is the same for all panels and represents the total amount of correlation between probes.}
\label{dcl}
\end{center}
\end{figure*}

We refer to \citet{hur15a} for a detailed modeling of tSZ and $\phi$ cross-power spectra, and to \citet{hur14} for the modeling of the X-ray emission.  
We relate cosmological parameters to the dark matter halo number per unit of mass and redshift, $\frac{{\rm d^2N}}{{\rm d}M {\rm d}V}$, using the mass function from \citet{tin08}.
\subsection{Poissonian term}
\label{1-halo}
Using the Limber approximation, we can write the one-halo term as
\begin{equation}
C_{\ell}^{\rm A\times B-{P}} = 4 \pi \int {\rm d}z \frac{{\rm d}V}{{\rm d}z {\rm d}\Omega}\int{\rm d}M \frac{{\rm d^2N}}{{\rm d}M {\rm d}V} W^{\rm P}_{\rm A} W^{\rm P}_{\rm B}.
\end{equation}
The tSZ contribution can be written as
\begin{align}
W^{\rm P}_{\rm tSZ} = Y_{500} y_{\ell},
\end{align}
where $Y_{500}$ represents the tSZ flux of the clusters within a radius where the matter density equals 500 times the critical density at the clusters' redshift, related to the mass contained in the same volume ($M_{500}$) via the scaling law presented in \citet{planckSZC}. In this same expression, $y_\ell$ represents the Fourier transform on the sphere of the cluster pressure profile per unit of tSZ flux from \citet{planckSZC} . We consider a GNFW Universal pressure profile \citep{arn10,planckppp}.

We model the lensing contribution as 
\begin{align}
W^{\rm P}_{\phi}~=~-2 \psi_\ell \frac{(\chi' - \chi) \chi}{\chi'},
\end{align}
with $\chi$ the comoving distance, $\chi'$ the comoving distance to the surface of the last scattering of the CMB, and $\psi_\ell$ the 3D lensing potential Fourier transform on the sky. We can express the potential $\psi$ as a function of the density contrast,
\begin{align}
\Delta \psi = \frac{3}{2} \Omega_m H_0^2 \frac{\delta_{\rm 3D}}{a},
\end{align}
with $a$ the universe scale factor and $\delta$ the density contrast. From this, the lensing contribution reads
\begin{align}
W^{\rm P}_{\phi} = \frac{3 \Omega_m H_0^2 (1+z)}{c^2 \ell (\ell+1)} \frac{(\chi' - \chi) \chi}{\chi'} \delta_\ell,
\end{align}
where  $\delta_{\ell}$ is the Fourier transform of the density contrast profile, $\delta_{\rm 3D}(u)$, computed as,
\begin{equation}
\delta_{\ell} = \frac{4 \pi r_{\rm 500}}{l^2_{\rm 500}} \int_0^{\infty} {\rm d}u \, u^2 \delta_{\rm 3D}(u) \frac{{\rm sin}(\ell u / \ell_{\rm 500})}{\ell u / \ell_{\rm 500}},
\end{equation}
where $u = r/r_{\rm 500}$ is the normalized radius of the profile, $\ell_{\rm 500} = D_{\rm A}/r_{500}$, $D_{\rm A}$ is the angular diameter distance, and $r_{500}$ is the radius within which the matter density is 500 times the critical density of the universe.\\

Finally, the X-ray contribution can be written as
\begin{align}
W^{\rm P}_{\rm X} = S_{500} x_\ell,
\end{align}
with ${S}_{500} = \overline{C} L_{500}$, the X-ray count-rate in the [0.5-2.0] keV energy band of the host halo, $L_{500}$ the unabsorbed X-ray luminosity in the [0.1-2.4] keV energy range, $\overline{C}$ the average luminosity to count-rate conversion factor described in \citet{hur14}, and $x_\ell$ the Fourier transform of the X-ray number count profile. To model the $L_{500}-M_{500}$ relation, we used the relation derived by \citet{arn10} from the \textsc{REXCESS} sample \citep{boh07}. We considered a polytropic equation of state with a polytropic index of 1.2 to compute the density and the temperature profiles from the pressure profile. \\

\subsection{Large-scale correlation terms}
\label{2halos}
We express the large scale correlations, the two-halo term, contribution as
\begin{align}
C_{\ell}^{\rm A\times B-C} = 4 \pi  \int {\rm d}z \frac{{\rm d}V}{{\rm d}z{\rm d}\Omega} W^{\rm C}_{\rm A} W^{\rm C}_{\rm B} P_k,
\end{align}
with $P_k$, the matter power-spectrum computed using {\tt CLASS} \citep{les11}.\\

For the tSZ effect, the CMB weak lensing, and the X-ray count rate we can express the window functions as,
\begin{align}
W^{\rm C}_{\rm tSZ} &= \int{\rm d}M \frac{{\rm d^2N}}{{\rm d}M {\rm d}V} Y_{500} y_{\ell} b_{\rm lin}, \nonumber \\
W^{\rm C}_{\phi} &= \frac{3 \Omega_m H_0^2}{c^2 \ell (\ell+1)} (1+z) \frac{(\chi' - \chi)}{\chi' \chi}. \nonumber \\
W^{\rm C}_{\rm X} &= \int{\rm d}M \frac{{\rm d^2N}}{{\rm d}M {\rm d}V} S_{500} x_{\ell} b_{\rm lin}.
\end{align}
where $b_{\rm lin}$ is the linear bias relating the halo distribution to the overdensity distribution. We considered the bias from \citet{mo96}, which is realistic on galaxy cluster scales. \\

We present in Fig.~\ref{dcl} the power density distribution in the $M_{500}$-$z$ plane for the (one-halo) X-ray, weak lensing, and tSZ auto- and cross-correlation power spectra at $\ell = 200$. We observe that weak-lensing favors higher redshift objects compared to the tSZ effect and X-ray emission. The X-ray and tSZ are highly correlated (at $\simeq 76\,$\%), while the tSZ effect and the weak lensing one-halo terms are moderately correlated  ($\simeq 46$\,\%), and the X-ray emission and the weak lensing one halo-terms show a lower correlation at the level of $\simeq $\,20\%. 
We refer the reader to \citet{hur15a} and \citet{hur15b} for a detailed description of the mass and redshift dependence of the CMB weak lensing and X-ray window functions. In the light of Fig.~\ref{dcl}, it turns clear that  detecting the X-ray--lensing cross-correlation is thus particularly challenging, considering the high-noise level of {\it Planck} CMB weak lensing maps, and the AGN-dominated X-ray sky.

\section{First measurement of the cross-correlation between X-rays and weak-lensing}
\label{secxphi} 

We use the ROSAT all-sky survey (RASS) public data\footnote{\url{ ftp://ftp.X-ray.mpe.mpg.de/rosat/archive/}}, which covers 99.8\% of the sky, including 97\% that has an exposure time longer than 100s \citep{vog99}. A description of the reprojection of the RASS data on the Healpix full-sky map can be found in \citet{hur15b}. We also use the {\it Planck} full-sky CMB weak lensing map \citep{planckphi}.
We mask all regions with less than 100s exposure time in RASS full-sky map, while using the mask associated to the {\it Planck} all-sky weak lensing map. This combination of masks results in an effective sky fraction of $f_{\rm sky} \simeq 75$\% in the cross-correlation. We have also verified that using more aggressive galactic masks (down to a sky coverage of 50\%) does not modify the results significantly.

\begin{figure}[!h]
\begin{center}
\includegraphics[width=0.9\linewidth]{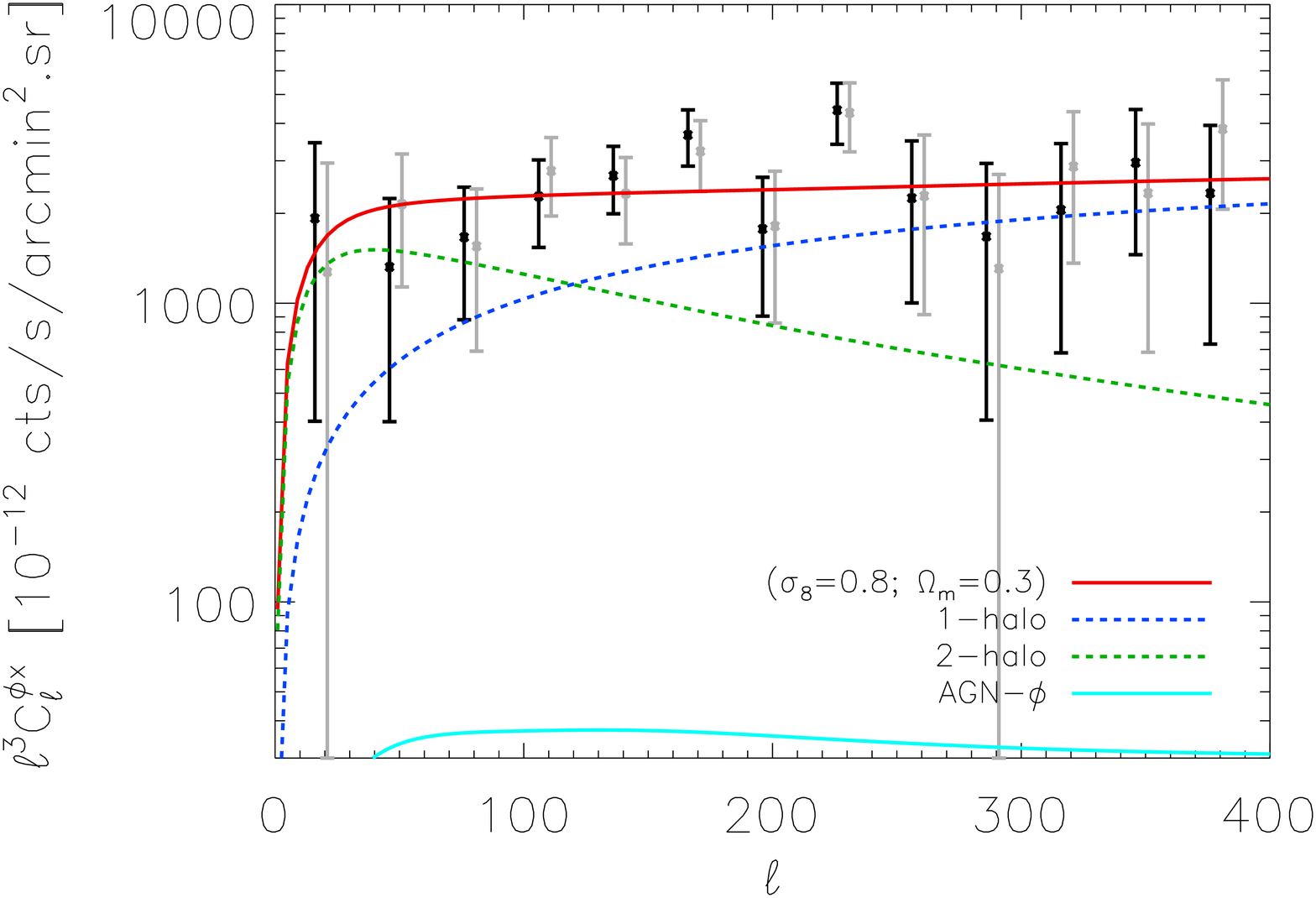}
\caption{X-ray-$\phi$ cross-correlation angular power spectrum, measured between {\it Planck} CMB lensing full-sky map and a RASS data reprojected on the full-sky (black sample), the same cross-spectra when masking NVSS sources is shown as grey samples. The red solid line shows the theoretical prediction assuming ($\sigma_8 = 0.8$, $\Omega_{\rm m} = 0.3$,  $b_H = 0.2$), the green dashed line shows the two-halo term contribution, and the blue dashed line the one halo term contribution. The solid cyan line displays the contribution from AGN-$\phi$ cross-correlation.}
\label{xphi}
\end{center}
\end{figure}

We compute the uncertainty in the weakly non-Gaussian limit as
\begin{align}
V(C_\ell^{x\phi}) = \frac{\left(C_\ell^{x\phi}\right)^2 + C_\ell^{xx}C_\ell^{\phi \phi}}{(2\ell +1)f_{\rm sky}},
\end{align}
where $C_\ell^{x\phi}$, $C_\ell^{xx}$, are $C_\ell^{\phi \phi}$ are the X-ray-lensing, X-ray, and weak lensing power spectra.
We obtained a significance of 9.1$\sigma$ for the X-ray-weak lensing cross-power spectrum in the range $\ell \in [0,400]$. 

We verified that AGNs do not produce a significant contribution to the total signal by masking BOSS AGNs \citep{alb16} and performing our analysis on the BOSS footprint ($f_{\rm sky} = 25$\%). We also masked NVSS bright sources, $S > 0.03$ Jy \citep{con98} and restrict our analysis to the NVSS footprint ($\delta > -40^{\rm o}$). We do not observe any significant modification of our results when masking the BOSS AGNs or NVSS sources. We also modeled the AGN contribution following the AGN mass and luminosity functions from \citet{hut14}. This contribution is about two orders of magnitude smaller than the X-ray-$\phi$ cross-correlation, as shown on Fig.~\ref{xphi}. Consequently, AGN impact on our results is small compared to the uncertainty level. We thus neglect the AGN contribution to the total X-ray-lensing cross-correlation. 

In Fig.~\ref{xphi}, we present the derived angular cross-power spectrum, compared to our modeling for ($\sigma_8 = 0.8$, $\Omega_{\rm m} = 0.3$,  $b_H = 0.2$). We observe that the two halo term dominates for $\ell < 100$ and that the one halo term dominates at higher multipoles. This illustrates how {\it Planck} weak lensing map contains a significant signal produced by compact objects like galaxy clusters.
Assuming a Gaussian prior of $b_H = 0.20 \pm 0.05$, consistent with hydrodynamical simulations, we derive $\Sigma_8 = \sigma_8  \left(\Omega_{\rm m}/0.30\right)^{0.27} = 0.80 \pm 0.03$.

\section{Combined analysis of thermal Sunyaev-Zel'dovich effect, X-ray, and weak-lensing signals}
\label{secdis}

We next combine our results with tSZ--$\phi$, tSZ--tSZ, tSZ--X-ray, $\phi$--$\phi$, and CMB--CMB power spectra results.
We assume that the CMB angular power spectrum is uncorrelated with all other probes\footnote{This is a reasonable approximation provided the low level of correlation induced by the integrated Sachs-Wolfe effect, which is restricted to very low multipoles ($\ell < 50$). We are also assuming that residual AGN and galaxy cluster tSZ residual contamination on the CMB should be of relevance at much higher multipoles than those considered here ($\ell > 1000$)}, so we compute the covariance between all large-scale structure tracers in the weakly non-Gaussian limit as,
\begin{align}
{\rm COV}(C_\ell^{\rm AB},C_\ell^{\rm CD}) = \frac{C_\ell^{\rm AC}C_\ell^{\rm BD} + C_\ell^{\rm AD}C_\ell^{\rm BC}}{(2\ell +1)f_{\rm sky}},
\end{align}
where A, B, C and D stand for the tSZ effect, the CMB weak lensing signal, and the X-ray emission. We use the tSZ analysis from \citet{hur17a}, the tSZ--$\phi$ measurement corrected for cosmic infra-red background contamination from \citet{hur15a}, the tSZ--X-ray results from \citet{hur15b}, the {\it Planck} weak lensing results from \citet{planckphi}, and the {\it Planck} CMB results from \citet{planckcmb}.

\begin{figure}[!h]
\begin{center}
\includegraphics[width=0.9\linewidth]{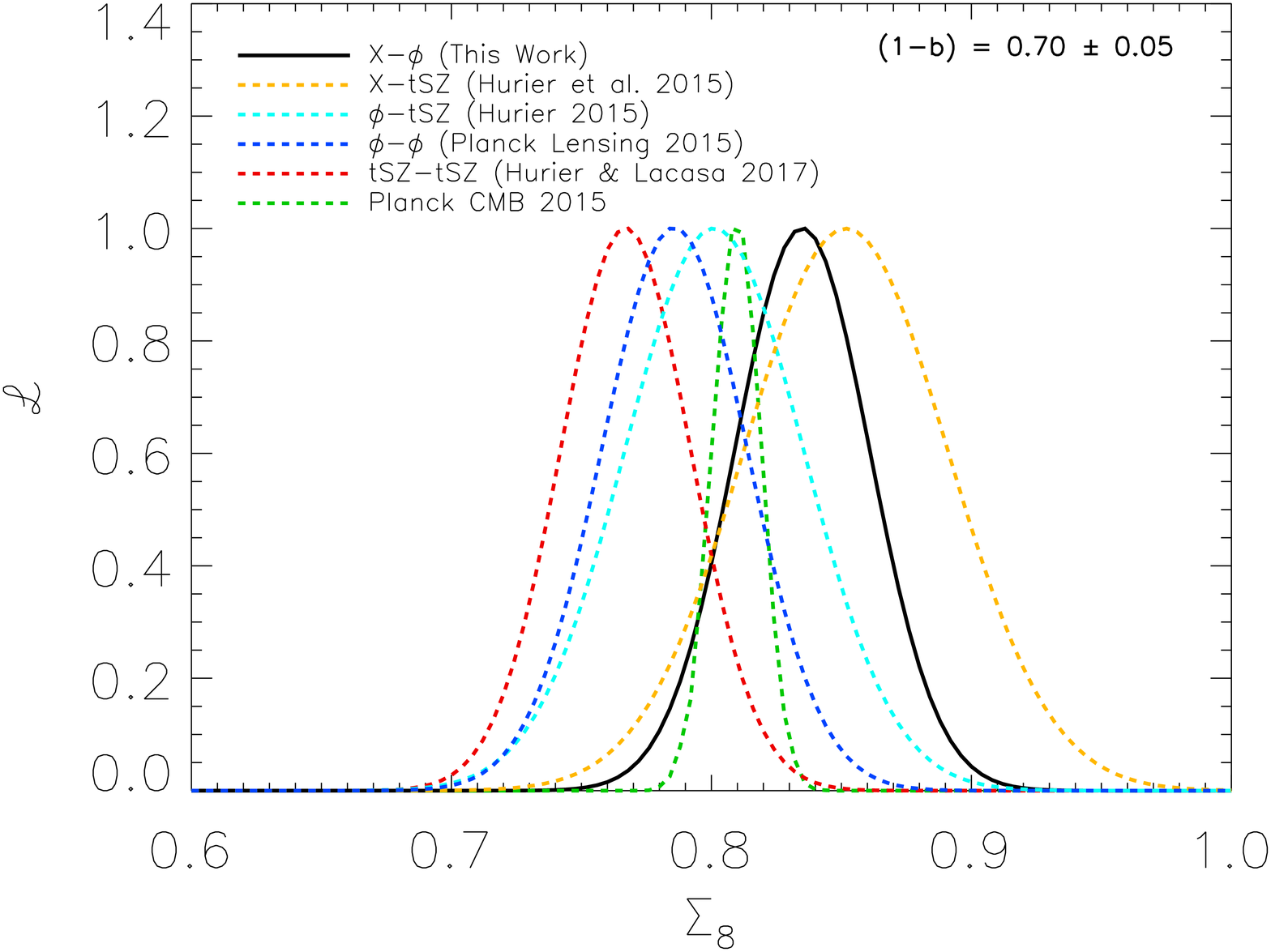}
\caption{Likelihood function of $\Sigma_8 = \sigma_8 \left(\Omega_{\rm m}/0.30\right)^{0.27}$ from different analyses: tSZ angular power spectrum, bispectrum, and number count \citep[red,][]{hur17a}, CMB weak lensing \citep[dark blue,][]{planckphi}, CMB angular power spectrum \citep[green,][]{planckcmb}, tSZ--weak lensing cross correlation \citep[cyan,][]{hur15a}, tSZ--X-ray cross-correlation \citep[orange,][]{hur15b}, and X-ray--$\phi$ cross-correlation (black, this work)}
\label{like}
\end{center}
\end{figure}

We performed a join fit of these results to derived cosmological constraints on $\sigma_8$ and $\Omega_{\rm m}$ and constraints on the hydrostatic mass bias, $b_H$. Both weak-lensing and CMB results are not sensitive to the hydrostatic mass bias. Consequently, weak lensing and CMB constraints set the cosmological parameters: $\sigma_8$ and $\Omega_{\rm m}$, whereas the large scale structure tracers, namely the tSZ effect and the X-ray emission, set the hydrostatic mass-bias value. From this combined analysis, we derived $b_H = 0.30 \pm 0.05$. 
We present the resultant likelihoods for tSZ, tSZ--X, tSZ--$\phi$, X--$\phi$, $\phi$, and CMB analyses for $b_H = 0.30 \pm 0.05$ in Fig.~\ref{like}. The X-ray auto-correlation power spectrum is not shown in this figure as it is particularly difficult to derive robust constraints from it, considering that the X-ray sky is dominated by AGN contribution.
We observe that some tension remains, especially with tSZ derived constraints, but all large scale structure analyses presented here are consistent within 2\,$\sigma$ with the {\it Planck} CMB results.

\section{Conclusion and discussion}
\label{seccon}

We have produced the first detection of the X-ray--$\phi$ cross-correlation angular power spectrum, with a significance of 9.1\,$\sigma$.
We have established cosmological constraints on $\sigma_8$ and $\Omega_m$ from this cross-correlation, that we find consistent with previous large-scale structure  \citep{hur15a,hur15b,hur17a,hur17c} and CMB analyses \citep{planckcmb,planckphi}.
Similarly to the tSZ--X cross-correlation, the X--$\phi$ correlation favors  values of the hydrostatic mass bias lower than those suggested in tSZ-CMB combined analyses \citep{sal17}. It also favors a higher value for $b_H$ than most of the weak-lensing based analyses of the last four years \citep[see e.g.,][]{med17,ser17,jim17,par17,oka16,bat16,app16,smi16,hoe15,sim15,isr15,don14,gru14,mah13}. This analyses prefer the range $b_H = 0.20 \pm 0.08$.
With the constraint inferred here ($b_H = 0.30 \pm 0.05$), large scale structure cosmological constraints from the local universe on $\sigma_8$ and $\Omega_m$ now surround the CMB-based cosmological constraints.
This result favors high value of $b$ compared to hydrodynamical simulations that prefer $b_H < 0.2$.
Additionally, under the assumption that these measurements are systematic-free, the significant difference between tSZ and X-ray based results may indicate that these two probes select significantly different populations of galaxy clusters in terms of hydrostatic mass bias.

\section*{Acknowledgment}
\thanks{We acknowledge the use of HEALPix \citep{gor05}}

\bibliographystyle{aa}
\bibliography{szlens_conta_cib,szphi_red}

\end{document}